\documentclass[twocolumn,aps,superscriptaddress,pra]{revtex4-1}

\usepackage{amsmath}
\usepackage{amssymb}
\usepackage{color} 
\usepackage{graphicx}
\usepackage[T1]{fontenc}
\usepackage[english]{babel}
\usepackage[normalem]{ulem} 

\def\Aarhus{Department of Physics and Astronomy, Aarhus University, DK-8000 Aarhus C, Denmark}

\def\PAS{Institute of Physics, Polish Academy of Sciences, Aleja Lotnikow 32/46, PL-02668 Warsaw, Poland}

\def\PKS{Max-Planck-Institut f\"{u}r Physik komplexer Systeme, D-01187 Dresden, Germany}

\begin{document}

\title{Global optimization for quantum dynamics of few-fermion systems}

\author{Xikun Li}
\affiliation{\Aarhus}

\author{Daniel P{\k e}cak}
\affiliation{\PAS}

\author{Tomasz Sowi{\'n}ski}
\affiliation{\PAS}

\author{Jacob Sherson}
\affiliation{\Aarhus}

\author{Anne E. B. Nielsen}
\altaffiliation{On leave from: \Aarhus}
\affiliation{\PKS}

\begin{abstract}
Quantum state preparation is vital to quantum computation and quantum information processing
tasks. In adiabatic state preparation, the target state is theoretically obtained with nearly perfect
fidelity if the control parameter is tuned slowly enough. As this, however, leads to
slow dynamics, it is often desirable to be able to do processes faster.
In this work,
we employ two global optimization methods to estimate the quantum speed limit for few-fermion
systems confined in a one-dimensional harmonic trap. Such systems can be produced experimentally
in a well controlled manner. We determine the optimized control fields and achieve a reduction in
the ramping time of more than a factor of four compared to linear ramping. We also investigate how
robust the fidelity is to small
variations of the control fields away from the optimized shapes.
\end{abstract}

\maketitle
\section{Introduction}
Quantum optimal control is essential to manipulate and engineer complex quantum systems in quantum information
processing and quantum computation~\cite{dalessandro2007,brif2010,krotov1996}. The operations in experiments
are often executed adiabatically to guarantee the transition to the target state with almost perfect
fidelity~\cite{gericke2007}. The adiabatic process, however, needs to be done slowly, and it is therefore interesting to look for ways to achieve a speed up which is the topic of the field of quantum optimal control~\cite{deng2018,modak2017}.

The minimal allowed time for driving such transitions with perfect fidelity is known as the quantum speed
limit (QSL)~\cite{mandelshtam1945,gajdacz2015}. The quantum speed limit is a lower bound for the duration
in which the quantum system can be completely steered to the target state~\cite{levitin2009,deffner2013,
campo2013,gajdacz2015,bason2012,jj2016}. For durations shorter than the quantum speed limit, defects emerge
that lead to a drop in fidelity between
the target state and the obtained state. The quantum optimal control
theory is important to obtain the quantum speed limit~\cite{lioyd2014} and has been applied using certain
numerical methods in many quantum systems like the NMR~\cite{glaser2015}, Bose-Einstein Condensates~\cite{doria2011,rosi2013,frank2016} and
spin chain models~\cite{burgarth2010,caneva2012}.

Except for a few special cases in which analytical results are available, one has to perform numerical calculations, which are highly non-trivial, due to the high dimensionality of the Hilbert space. Generally,
quantum control theory relies on numerical techniques including the local optimization algorithms, such as
Krotov, GRAPE and CRAB~\cite{caneva2011,sklarz2002,khaneja2005,machnes2005}, as well as the global optimization
methods like Differential Evolution (DE)~\cite{zahedinejad2014,palittapongarnpim2017,brest2006} and covariance matrix
adaptation evolution strategy (CMA-ES)~\cite{shir2011,hansen2006}. In Ref.~\cite{jj2018}, it is proposed that numerical
optimization relies on an appropriate balance between local and global optimization approaches and problem
representation. When the quantum system is fully controllable and free of constraints, there are no traps in
the form of suboptimal local extrema~\cite{rabitz2004}. In such cases, the local algorithms are preferred as
the computational cost of local optimization methods is lower than that of the global ones. When the duration
of the process is short or if there are constraints on the control field, the local algorithms are often stuck in the local
suboptimal traps in the quantum control landscape. For the low-dimensional quantum system, the computational
cost of multistarting the local optimization algorithms, which is able to give sufficiently good results, is
comparable with that of global ones. In Ref.~\cite{zahedinejad2014}, the local optimization methods fail to obtain a
satisfactory result determined by certain threshold infidelity for quantum gates, though the global optimization methods succeed.
The superiority of global optimization methods are also highlighted for
high-dimensional Hamiltonians studied in the Ref.~\cite{palittapongarnpim2017}.

The ultimate goal is to fully control any many-body quantum system. In cold atom experiments, one can influence the interparticle interaction with an external magnetic field thanks to Feshbach resonances. Due to the adiabatic change of the interaction it is possible to obtain for example a highly correlated state known as the Tonks-Girardeau gas starting from the non-interacting state~\cite{kinoshita2004observation,paredes2004tonks,haller2009realization}. Unfortunately, the full control of systems with a large number of particles is very challenging. A possible way to overcome the difficulty of such complex systems is to fully control smaller physical systems and use them to build the real many-body systems. A possible candidate to serve this purpose are quantum systems of a few ultra-cold atoms~\cite{serwane2011,zurn2012fermionization,zurn2013Pairing,wenz2013,Murmann2015AntiferroSpinChain,Murmann2015DoubleWell,Kaufman2015Entangling}. In the two-component mixtures of fermions one can deterministically prepare a system confining a well-established number of atoms with astonishing precision. The properties of few-body ultra-cold systems were also studied recently theoretically, including energy spectra and density profiles~\cite{Blume2013,Sowinski2015Pairing,lindgren2014,Grining2015,decamp2016,Deuretzbacher2014,
yang2015,yang2016,DAmico2015Pairing}. The two different flavors in the mixture of same-mass fermions are realized experimentally by using two different hyperfine states of ultra-cold lithium $^6$Li. A natural way to generalize this idea is to change the hyperfine states to two completely different species, for example lithium and potassium~\cite{tiecke2010Feshbach6Li40K,Wille6Li40K}. Such an experiment on a two-flavor mixture of lithium and potassium on a many-body scale has already been performed~\cite{cetina2016}. Recently, few-body mass-imbalanced systems were broadly explored theoretically~\cite{pecak2016,pecak2016b,pecak2017b,garcia2016,loft2015,harshman2017,volosniev2017}.

In this paper, we employ two global optimization algorithms, CMA-ES~\cite{shir2011,hansen2006} and
self-adaptive DE (SaDE)~\cite{brest2006,palittapongarnpim2017}, to numerically estimate the quantum
speed limit for few-fermion mass-imbalanced systems, and show the optimized control field for various durations.
We consider the fidelity of the final state with respect to the target state
as the fitness function to be optimized. As a proof of concept we show
how to fully control a system of a few fermions and drive it from the non-interacting state to the strongly correlated one.

\section{The model}\label{sec2}

We consider a two-flavor system of a few ultracold fermions confined in a one-dimensional harmonic trap.
Here we assume that the frequencies $\omega$ of the harmonic trap are the same for both flavors. Fermions
of opposite flavors interact in the ultra-cold regime via short-range forces modeled by the delta-like potential
$U(x-x') = g\delta(x-x')$, where $g$ is the interaction
strength~\cite{Olshanii1998}. In this approximation fermions of the same type do not interact as a consequence of the Pauli exclusion principle. Fermions of opposite flavors are fundamentally distinguishable and may have the same or
different masses (in the following we denote the mass ratio $\mu=m_\uparrow/m_\downarrow$). The Hamiltonian
of the mass-imbalanced system reads (see \cite{pecak2016,pecak2016b,pecak2017b}):
\begin{align} \label{eq:hamiltonian}
 \hat{\cal H} &= {\sum_{i=1}^{N_{\downarrow}} \left[ - \frac{1}{2}
 \frac{\partial^2}{\partial x_i^2} + \frac{1}{2} x_i^2  \right]
+ \sum_{{j}=1}^{N_{\uparrow}}} { \left[ - \frac{1}{2 \mu} \frac{\partial^2}
{\partial y_{{j}}^2} + \frac{\mu}{2} y_j^2 \right] } \nonumber\\
&+ g \sum_{i,j=1}^{N_{\downarrow},N_{\uparrow}} \delta(x_i - y_j).
\end{align}
All quantities are measured in appropriate harmonic oscillator units, i.e., positions are measured in units of $\sqrt{\hbar/(m_\downarrow \omega)}$, time in units of $1/\omega$, energies in $\hbar \omega$, and the interaction strength $g$ is measured in units of $(\hbar^3 \omega/ m_\downarrow)^{1/2}$.

\begin{figure}
\includegraphics[width=0.71\linewidth]{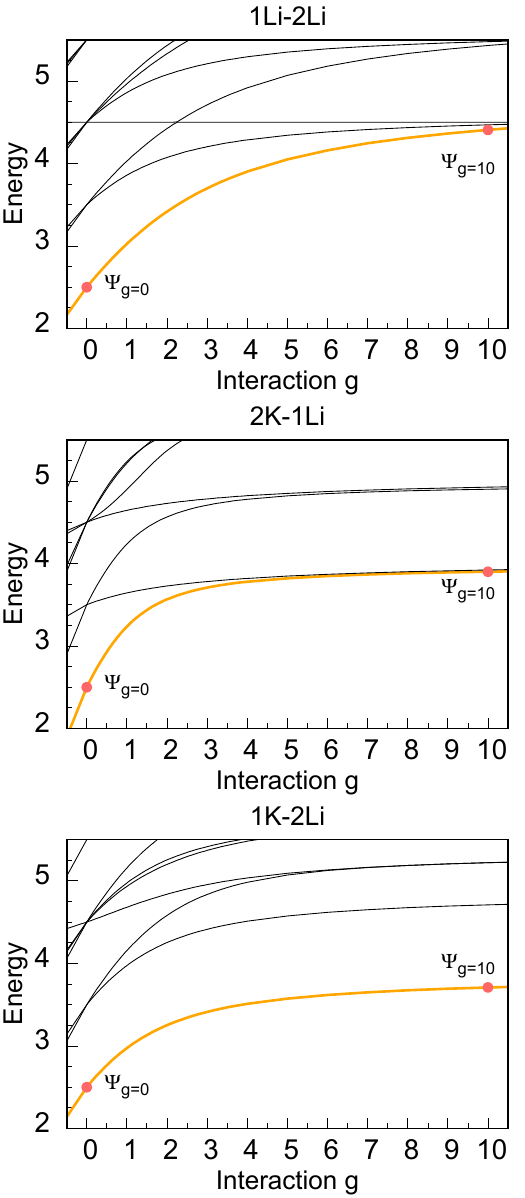}
\caption{The energy spectrum of the system of three fermions for three different scenarios, namely 1Li-2Li, 2K-1Li, and 1K-2Li. The thick orange line gives the ground-state energy. The non-interacting ground state $g=0$ and the strongly correlated target state $g=10$ are marked with the red dots. For the 1Li-2Li and the 2K-1Li systems, there are, respectively, 2 and 1 other states that have energies close to the ground state energy in the strong coupling limit. These states do, however, have different symmetries than the ground state, and the population of these states remain zero throughout the dynamics. The energies and the interaction strengths are measured in units of $\hbar\omega$ and $(\hbar^3\omega/m_\downarrow)^{1/2}$, respectively.
}
\label{fig:spectrum}
\end{figure}

We employ the exact diagonalization approach to study the dynamics of the few-fermion system (see Ref.~\cite{pecak2016}
for details of the numerical method).
The interaction $g$ is controlled experimentally with the help of the
magnetic field $B$, so by changing the magnetic field
in time, one can also modify the interaction $g(B(t))$.
Thus, it is convenient to treat the interaction strength $g(t)$ as the control field.
The many-body spectrum for the system of three fermions is shown in Fig.~\ref{fig:spectrum}. In this paper, we focus on
the transformation from the ground state of the non-interacting Hamiltonian $\hat{\cal H}(g=0)$ to that of
$\hat{\cal H}(g=10)$, where strong correlations are present. By increasing the interaction adiabatically, one can transfer the non-interacting state $\Psi_{g=0}$ to the interacting one $\Psi_{g=10}$. Note that depending on whether the mass-imbalance is present in the system, and on the specific configuration, the ground state of the system might be quasi-degenerated in the strong interaction limit. For the equal-mass system there is a three-fold degeneracy, while for the system 1K-2Li the ground-state is not degenerated. For the dynamics considered below, there is no coupling to these additional low-energy states, because they have a different symmetry. The time scale for adiabatic ramping is hence not determined by the energy of these states relative to the ground state, but by the energy of the higher lying states relative to the ground state.

For a typical quantum control problem, the Hamiltonian depends on a time-dependent control field
${\cal H}={\cal H} (g(t))$. We wish to optimize the fidelity as a fitness function
\begin{align}
F(g(t),T)&=|\langle\Psi_{g=10}|\mathcal{T}\exp (- \mathrm{i}\int_0^T {\cal H} \left(g(t)) \mathrm{d}t
\right) |\Psi_{g=0}\rangle|^2
\end{align}
with the time evolution driven by the control field $g(t)$, where $\mathcal{T}$ is the time-ordering
operator and the duration $T$ is discretized with time-step size $\delta t=10^{-3}$ for numerical evaluation of
the time evolution. It is worth noting
that the most time-consuming part during the numerical calculation is the time evolution which consists
of a long sequence of evaluations of exponentials of large Hamiltonian matrices (see Sec.~\ref{sec5}). Therefore
the maximal number of iterations is set to be 500 for CMA-ES and SaDE.

To perform this
optimization systematically, we choose $g(t)$ to be decomposed into a truncated Fourier basis (CRAB method, see
\cite{doria2011,caneva2011})
\begin{equation} \label{CRAB}
g(t) = g_0(t)\left[1 + \frac{1}{{\cal N} (t)}\sum_{n=1}^
{N_c} (A_n \cos(\omega_n t)+B_n \cos(\omega_n t))\right],
\end{equation}
where $g_0(t)$ is the initial guess of the control field and ${\cal N}(t)=T^2/2t(T-t)$ is a time-dependent function
to fix the initial and final control field value to be $g(t=0)=0$ and $g(t=T)=10$. $\{A_n, B_n\}$ are
Fourier coefficients and $\omega_n=2\pi n(1+r_n)/T$ are ``randomized'' Fourier harmonics, and $r_n\in [0,1]$.
The choice of the cut-off number $N_c$ of the Fourier basis may vary from
one case to another: it may depend on the Hamiltonian, the fitness function and the optimization algorithm. The parameter
space (search space) consists of the Fourier coefficients and harmonics $\{A_n, B_n, \omega_n\}$, which can be numerically
obtained using an optimization method, e.g., the simplex method, gradient-based strategies and global
optimization algorithms. In this paper, we restrict to non-negative interactions $g(t)\geq 0$ by simply putting
the negative values of $g(t)$ to be zero, in which
case the local optima in the quantum control landscape are usually not global optima.

\section{Global optimization}\label{sec3}
We employ two evolutionary computation techniques, CMA-ES and SaDE, as global optimization methods. We compare CMA-ES with SaDE for three different systems: (\emph{i}) the mixture of three $^6\mathrm{Li}$ atoms with two different hyperfine states ($N_\uparrow=1$, $N_\downarrow=2$, $\mu=1$); (\emph{ii}) the mixture of one $^{40}\mathrm{K}$ atom and two $^6\mathrm{Li}$ atoms ($N_\uparrow=1$, $N_\downarrow=2$, $\mu=40/6$); (\emph{iii}) the mixture of two $^{40}\mathrm{K}$ atoms and one $^6\mathrm{Li}$ atom ($N_\uparrow=2$, $N_\downarrow=1$, $\mu=40/6$). We numerically estimate the duration $T_{\mathrm{QSL}}$ for which the fidelity $F(T_{\mathrm{QSL}})=0.99$. We then compare the control field $g(t)$ for various durations $T$ and depict the deviations between the optimized control field and non-optimal ones. For simplicity, we will present results on the 1K-2Li system unless stated otherwise.

CMA-ES and SaDE are variants of evolution strategy (ES) and DE, respectively. Both ES and DE belong to the class of evolutionary algorithms and are stochastic, derivative-free algorithms for global optimization of fitness functions. An evolutionary algorithm works through a loop of variations (including recombination and mutation) and selection in each iteration (also called generation). New candidates are generated by variation of current parent individuals in each iteration. Then some candidates are selected, based on their fitness, to be parents for the next generation. In this way, search points with better and better values of the fitness function are generated over the sequence of iterations.

\begin{table}[!htb]
\caption{\label{table1}
Fidelity for various combinations of cut-off number $N_c$ and population size $N_p$ with duration $T=0.1$ using the CMA-ES method for the 1K-2Li system. The maximal value of fidelity is indicated by the bold font, and the corresponding pair of values is $N_c=15$ and $N_p=60$, respectively.}
\centering
\begin{tabular}{c|cccc}\hline \hline
$\mathbf{N_c}$& 5 & 10 & 15 & 20 \\ \hline
$\mathbf{N_p}$&&&& \\
20 & 0.5995 & 0.5869 & 0.5933 & 0.5857 \\
40 & 0.6021 & 0.5984 & 0.5924 & 0.5848 \\
60 & 0.5968 & 0.6015 & \textbf{0.6053} & 0.5940 \\
80 & 0.5963 & 0.5972 & 0.5987 & 0.5894 \\ \hline \hline
\end{tabular}
\end{table}
In CMA-ES, new search points (parameter vectors) are sampled according to a multivariate normal distribution in the parameter space. CMA-ES begins with a randomly initiated population of search points in the parameter space with initial mean and covariant matrix. The population size $N_p$ is the number of search points in each iteration. In the selection and recombination step, the search points with the best $m$ fitness, where $m$ is the parent size and not larger than the population size,  are chosen as the parents to update the new mean, step-size and covariant matrix. Recombination amounts to selecting a new mean value for the multivariate normal distribution. In the mutation step, the parameter vectors are further added by random vectors with zero mean and updated covariance matrix. The fitness function evolves iteratively towards its optimal state. In contrast to most other evolutionary algorithms, CMA-ES is quasi parameter-free: one needs only to randomly choose an initial value of the step-size. In addition, the population size $N_p$ does not depend on the dimension of the parameter space and can hence be chosen freely (which is in contrast to DE). In general, large population sizes help to circumvent local optima, while small population sizes usually lead to faster convergence.
Therefore, a trade-off between the computational cost and performance needs to be carefully determined if the computational time for each iteration is considerably long. See~\cite{hansen2006} for a review of the CMA-ES method.

In Table~\ref{table1} we show the values of fidelity for various combinations of $N_c$ and $N_p$ with duration $T=0.1$ obtained using the CMA-ES method for the 1K-2Li system. The maximal fidelity in Table~\ref{table1} is $F=0.60537$ with ($N_c=15, N_p=60$).  It is hardly possible to infer the optimal combination ($N_c, N_p$) to obtain the maximal fidelity for arbitrary durations, nor reasonable to try all possible combinations of ($N_c, N_p$) as the computation cost is huge. Therefore we fix the population size to $N_p^{\mathrm{ES}}=60$ and the cut-off number of the Fourier basis $N_c=15$ for all durations in the three different few-fermion systems.

In SaDE, an initial population of parameter vectors (called genome or chromosome) is randomly sampled. Then the mutant chromosomes are obtained from the differential mutation operation (origin of the term ``DE''). In the mutation step, three mutually exclusive parameter vectors are generated randomly. The new set of parameter vectors are generated by adding one of those three vectors to the difference between the other two vectors with the mutation scale factor $S$ which controls the differential variation. In the recombination step, an offspring is formed by recombining the original and those mutant chromosome in a stochastic way, where the crossover rate $Cr$ controls the probability of recombination. In the selection step, comparison of values of the fitness function determines whether the offspring or the original chromosome survives to the next generation. See Ref.~\cite{das2011} for a review of DE.

In the conventional DE algorithm, there are two free parameters: $S$, $Cr$ which are fixed through the iterations. In SaDE, however, $S$ and $Cr$ are adapted in each iteration to enhance the convergence rate for the high-dimensional optimization problem, and to obtain better quality solutions more efficiently, compared with the conventional DE. A reasonable value of $N_p$ for DE and its variants is usually chosen between $5D$ and $10D$ ($D=3 N_c$ is the dimension of the parameter space), as suggested in the field of evolutionary computation science~\cite{storn1997}. Note, however, that this is not tested in great detail for physically motivated quantum systems. In this work, we fix the population size $N_p=5D$ and set the cut-off number to $N_c=5$, thus $N_p^{\mathrm{DE}}=75$, for SaDE to reduce the computational cost. Details of SaDE can be found in~\cite{brest2006,palittapongarnpim2017}.

\section{Results}\label{sec4}

\begin{figure}
\includegraphics[width=7cm]{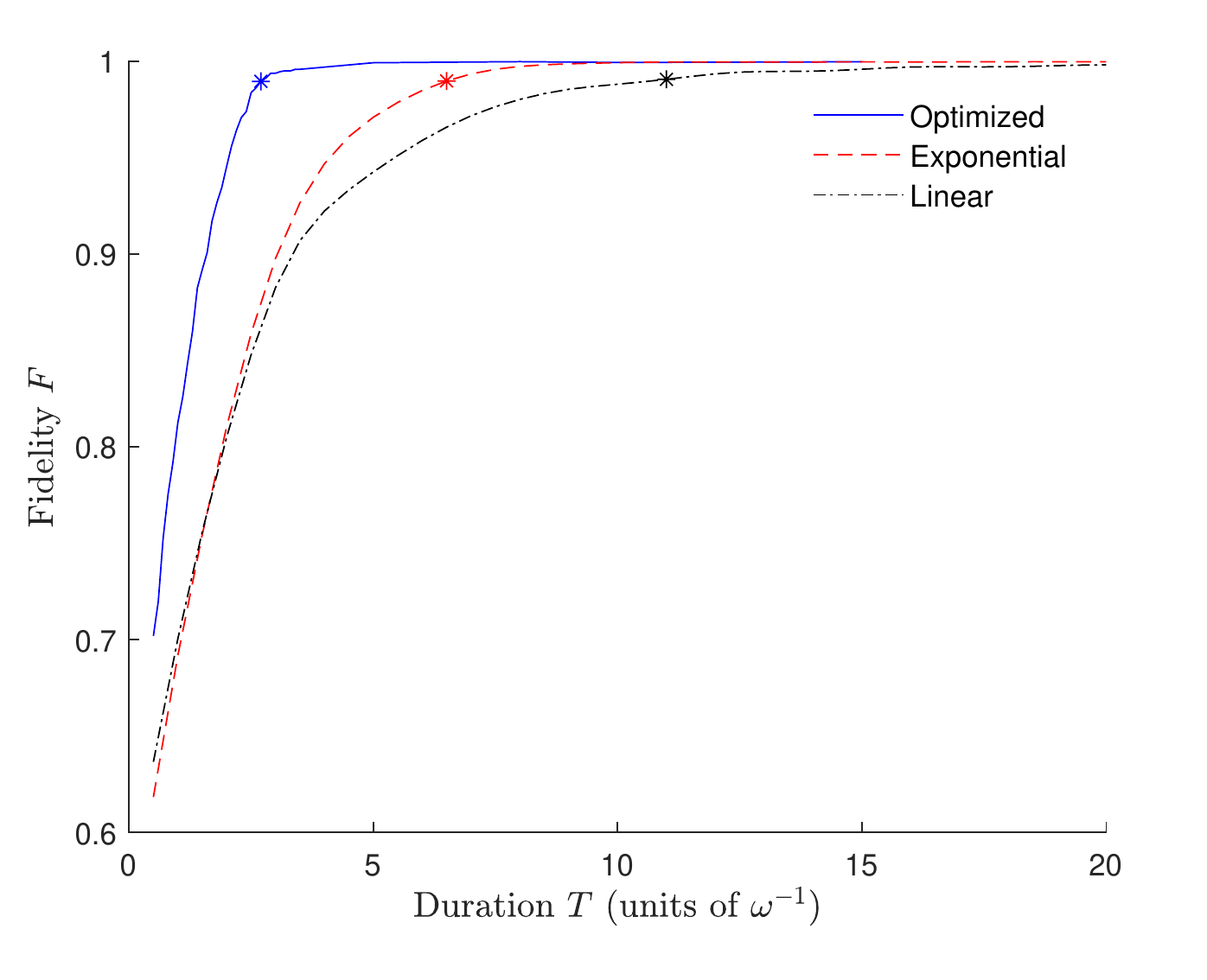}
\caption{Fidelity versus duration for the 1K-2Li system using different control fields: optimized (blue, solid line), exponential (red, dashed) and linear (black, dash-dot). The curve for the optimized control field is obtained using the CMA-ES algorithm. The stars mark the shortest durations for which a fidelity of $F=0.99$ is reached: $T_{\mathrm{Opt}}=2.7$, $T_{\mathrm{Exp}}=6.5$ and $T_{\mathrm{Lin}}=11$ (the time unit is $\omega^{-1}$).}
 \label{fig:QSL_elo}
\end{figure}

First, we numerically compute the fidelity obtained for optimized and non-optimal control fields, when the duration $T$ of the time evolution is fixed to a certain value. We use the constraint that the interactions must be nonnegative at all times. We consider two typical non-optimal rampings: linear ramping and exponential ramping. The latter is described by
\begin{equation} \label{expo}
g(t) = g_{\mathrm{max}} \frac{1-e^{t/\tau}}{1-e^{T/\tau}},
\end{equation}
where $\tau=T/5$ (In~\cite{rosi2013}, the exponential ramping with particular values of $T$ and $\tau$ is representative of quasi-adiabatic ramping of the lattice depth used in experiments of optical lattices). In Fig.~\ref{fig:QSL_elo}, we take the 1K-2Li system and compare the fidelity (as a function of duration) for the optimized ramping obtained using the CMA-ES method with that for exponential ramping and linear ramping. The shortest duration with fidelity $F=0.99$ is $T_{\mathrm{Exp}}=6.5$ for the exponential ramping and $T_{\mathrm{Lin}}=11$ for the linear ramping, whereas $T_{\mathrm{Opt}}=2.7$ for the optimized ramping (note the time unit is $1/\omega$). Thus the shortest duration obtained by the optimized control is approximately one fourth of that using linear ramping, and two fifth of that using the exponential ramping, in the 1K-2Li system. It means that by controlling the interaction in the optimized way, one can significantly reduce the time of preparing the
system in the strongly correlated state.

\begin{figure}
\includegraphics[width=1\linewidth]{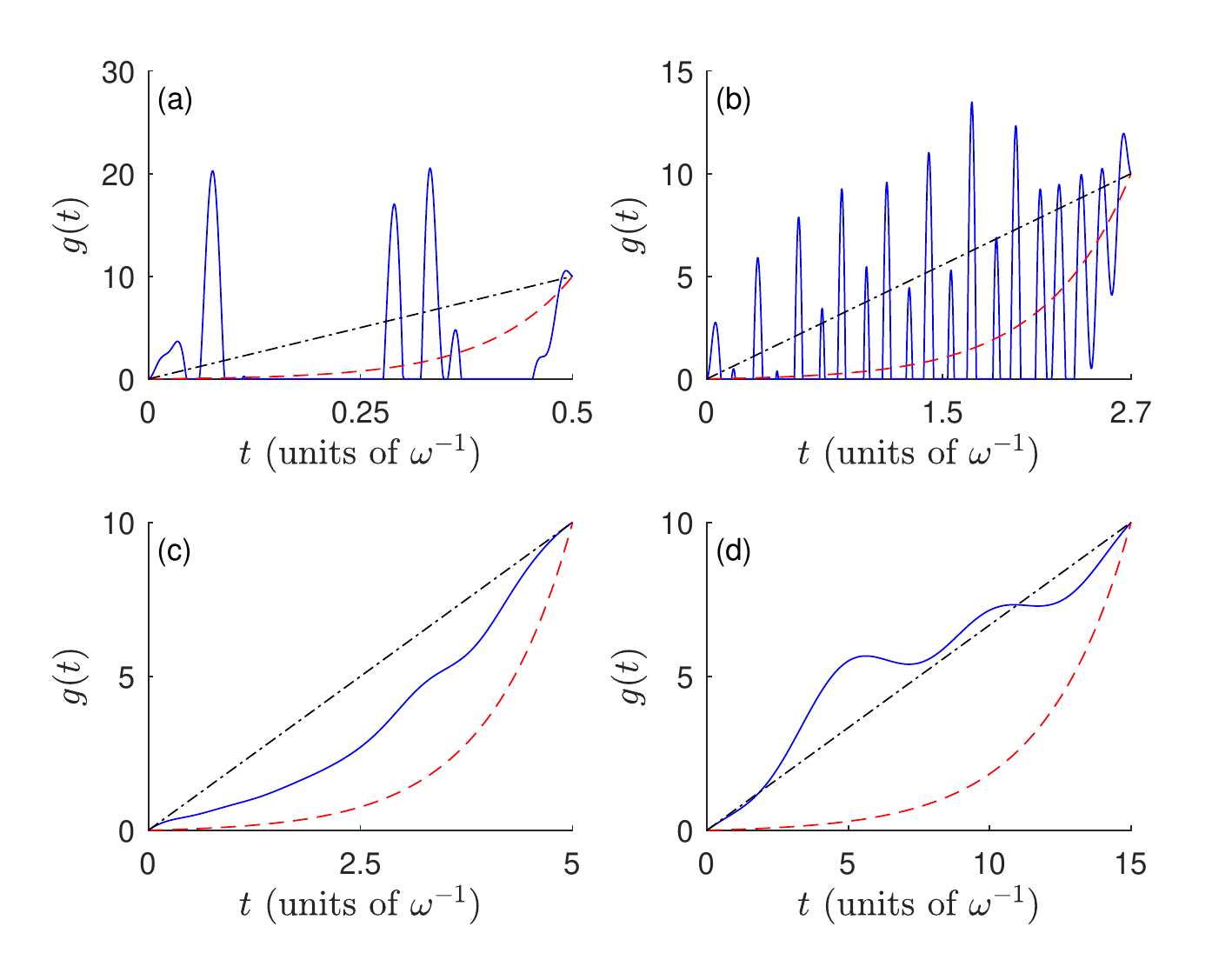}
\caption{Control field for different durations in the 1K-2Li system. (a) $T=0.5$, (b) $T=2.7$, (c) $T=5$ and (d) $T=15$ (the time unit is $\omega^{-1}$). The blue solid lines are the optimized control fields obtained using the CMA-ES method, the red dashed lines are exponential rampings, and the black dash-dot lines are linear rampings.}
 \label{fig:control}
\end{figure}

In Fig.~\ref{fig:control}, we show the optimized control field, the control field of linear ramping and exponential ramping for different durations. For very short duration ($T=0.5$), the optimized control field shows a few large oscillations, while $g(t)$ remains to be zero for most of the time (Note the requirement that $g(t)\geq 0$ ). As the duration approaches $T_{\mathrm{Opt}}=2.7$, which is an estimate of the quantum speed limit $T\approx T_{\mathrm{QSL}}$ (Fig.~\ref{fig:control}b), more oscillations emerge to make the transformation as fast as possible and the fidelity as large as possible. The deviation between the optimized control field and the linear and exponential ones is thus large for such durations. When the duration is much larger than $T_{\mathrm{Opt}}$, e.g., $T=5$ (Fig.~\ref{fig:control}c) and $T=15$ (Fig.~\ref{fig:control}d), the highly oscillating components are no longer necessary and the deviations between different rampings are much smaller than the cases in which $T<T_{\mathrm{Opt}}$. Also the fidelities do not depend for
the longer durations that much on the way we approach the strong interaction. This observation means that for long enough times, the exact shape of the control field
does not matter. This is why slow enough processes are quasi-adiabatic and the quantum state can be transferred when carefully managed. Since we require the interaction strength to be non-negative $g\geq0$, the lower bound of the control field is zero, but there is no upper bound. As mentioned above, if $g(t)$ (as given in Eq.~(\ref{CRAB})) is negative in some time interval, we put $g(t)$ equal to zero in that interval, as shown in Fig.~\ref{fig:control}. Such processes introduce the possibility of sharp peaks in the control fields and thereby high Fourier components. Given the particular experimental constraints of the system, one may perform an analogue optimization with appropriate constraints included.

\begin{figure}
\includegraphics[width=1\linewidth]{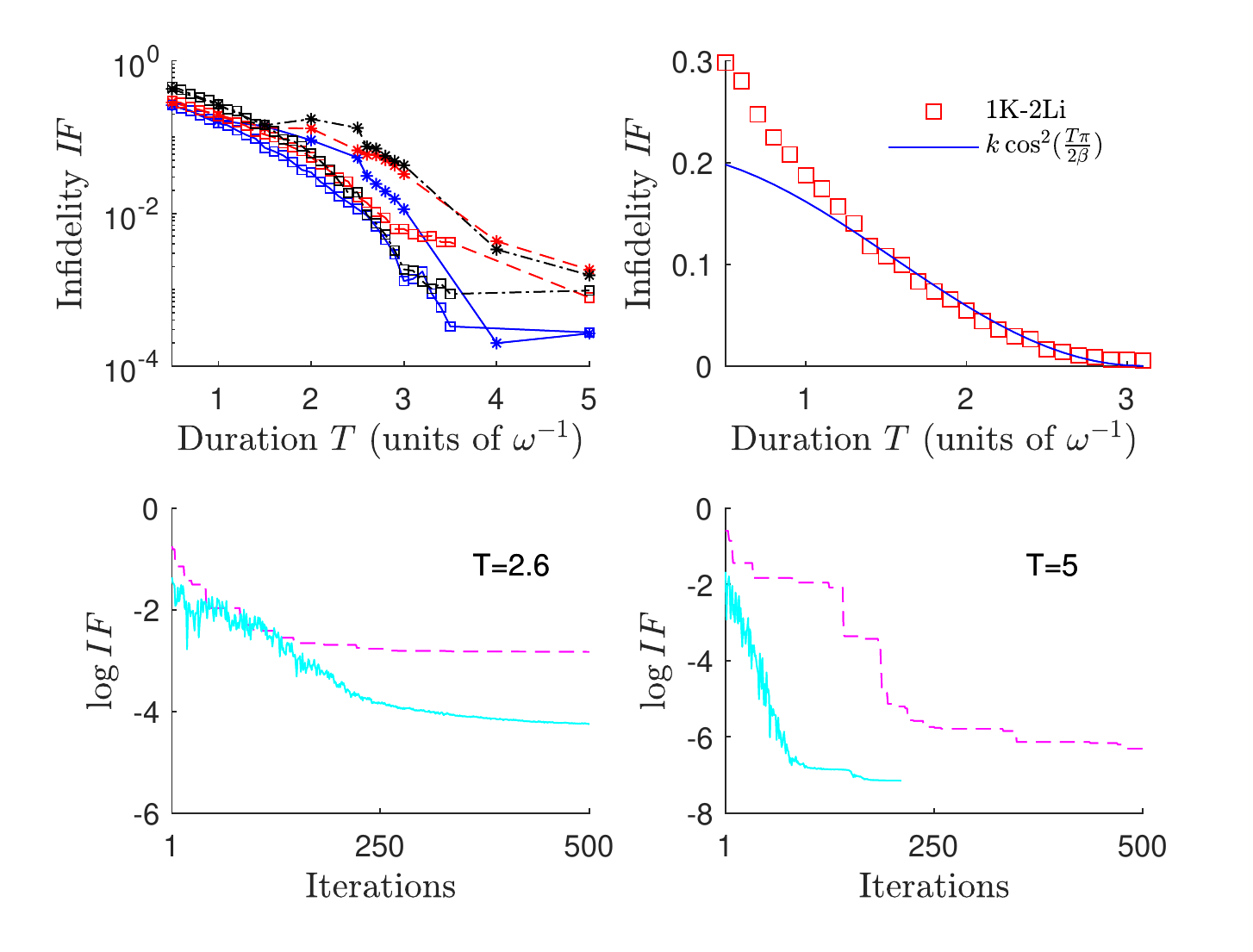}
\caption{Comparing CMA-ES and SaDE. The top left panel shows the infidelity ($I\!F=1-F$) as a function the duration $T$ using CMA-ES (square) and SaDE (asterisk) for 1Li-2Li (blue, solid), 1K-2Li (red, dashed) and 2K-1Li (black, dash-dot), where the scale of vertical axis is logarithm. In the top right panel, the red squares are numerical data, which are obtained using CMA-ES method, of infidelity $I\!F$ as a function of duration $T$ for the 1K-2Li system, while the blue solid lines are the fitting obtained using the function $k \cos^2(\frac{T\pi}{2\beta})$ for the numerical results, where $k$ and $\beta$ are free fitting parameters. In the bottom panel, the comparisons of $\log$-infidelity versus iterations in the 1K-2Li system are shown between CMA-ES (cyan, solid) and SaDE (magenta, dashed) for durations $T=2.6$ (bottom left) and $T=5$ (bottom right) with the time unit $\omega^{-1}$.}
 \label{fig:DE-ES}
\end{figure}

The comparisons between CMA-ES and SaDE for three different few-fermion systems are shown in Fig.~\ref{fig:DE-ES}. To reduce the computational cost, we set the maximal number of iterations to 500. In addition, we set up halt criteria for the CMA-ES and SaDE methods. For both of them, the calculations stop if the distance between the minimal and the maximal fidelity in the population is smaller than a threshold value ($Error=10^{-6}$). The top left panel depicts the infidelity $I\!F=1-F$ as a function of duration $T$ for three different systems using the CMA-ES and SaDE methods. For short durations $T<1.5$, the infidelities obtained using CMA-ES and SaDE are very close. For long durations $T>1.5$, however, the infidelities obtained by CMA-ES are smaller than SaDE (apart from $T=4$ for the 1Li-2Li system), which means the performance of CMA-ES is better than SaDE in terms of the best infidelity with the specific parameters ($N_p, N_c$) used for both global optimization methods in this work.
It is worth noting that the estimate of QSL is approximately proportional to the inverse of energy gap, i.e., $T_{\mathrm{QSL}}\sim \pi/\Delta$, where $\Delta\approx1$ is the energy gap to the nearest coupled excited state for all three few-fermion systems (See Fig.~\ref{fig:spectrum}). This fact agrees with the conclusion obtained in Ref.~\cite{caneva2011}. In the top right panel, we show the numerical results of infidelity $I\!F$ as a function of duration $T$ for the 1K-2Li system using the CMA-ES method and the curve fitting using the cosine square function $k \cos^2(\frac{T\pi}{2\beta})$ with two free fitting parameters $(k, \beta)$. The single cosine square function does not fit well for the numerical results. Such deviations or discrepancies are also found in different quantum systems~\cite{frank2016,jj2016,marin2017}.

In the bottom panel of Fig.~\ref{fig:DE-ES}, we demonstrate the $\log$-infidelity versus iterations (also called generation in evolutionary computation) for duration $T=2.6$ (bottom left) and $T=5$ (bottom right) in the 1K-2Li system. We observe that the $\log$-infidelity starts converging after a certain number of iterations (several tens to several hundreds) for the CMA-ES method, while the ``staircase'' pattern emerges in most cases of the SaDE method. From the lower panel in Fig.~\ref{fig:DE-ES}, the CMA-ES method performs better than the SaDE method in terms of the best infidelity and convergence rate. The reason why CMA-ES performs better than SaDE might be that the cut-off number of the Fourier basis of CMA-ES $N_c^{\mathrm{ES}}=15$ is larger than that of SaDE $N_c^{\mathrm{DE}}=5$ , such that the search space of CMA-ES is larger than that of SaDE. Note, however, the population size of CMA-ES ($N_p^{\mathrm{ES}}=60$) is smaller than that of SaDE ($N_p^{\mathrm{DE}}=75$).

Since the numerical calculations studied in this work are considerably time-consuming, the convergence rate of the optimization method is one of the most important considerations. Therefore, CMA-ES is more preferable than SaDE because CMA-ES does not need to use a large population size~\cite{hansen2006}. When the computational cost is small, the primary consideration is whether a satisfactory result, e.g., a certain threshold value of the fitness function, is achieved by the optimization method~\cite{zahedinejad2014}.

\begin{figure}
\includegraphics[width=1\linewidth]{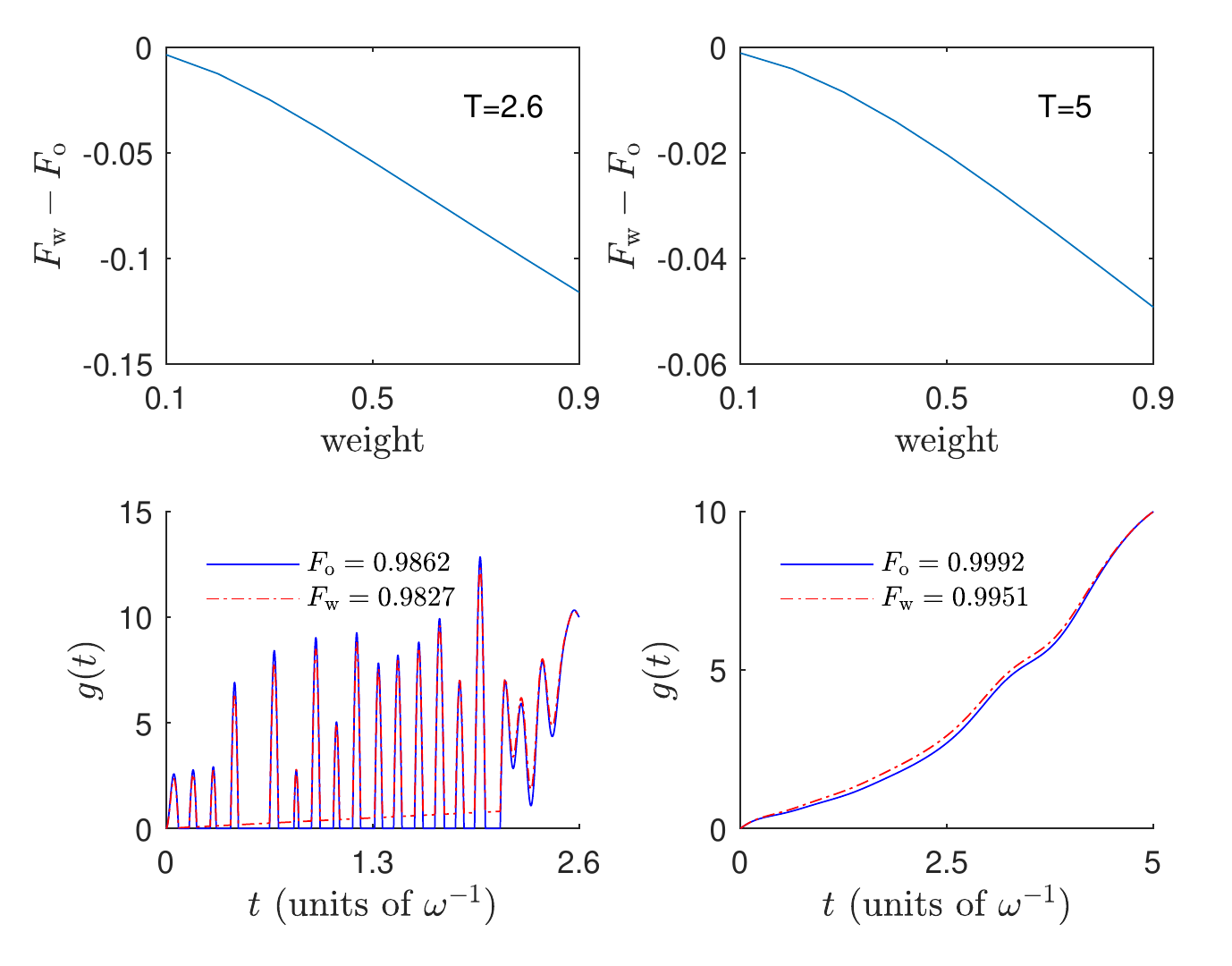}
\caption{Effects of deviations from the optimized control fields, which is obtained mixing the optimized control field and linear ramping with certain weight. In the top panel, the deviation between the fidelity of the modified optimized control field $F_\mathrm{w}$ and that of the optimized control field $F_\mathrm{o}$ are shown as a function of weight for $T=2.6$ (left) and $T=5$ (right) with the time unit $\omega^{-1}$. In the bottom panel, both the optimized control field (blue, solid line) and its modification whose weight is $10\%$ (red, dash-dot) are shown for $T=2.6$ (left) and $T=5$ (right). The fidelity of the optimized control field $F_\mathrm{o}$ and that of the modified one $F_\mathrm{w}$ are also given, and the difference between them is of order $10^{-3}$ for both durations.}
 \label{fig:opt-var}
\end{figure}

From the experimental point of view a very important question arises: how sensitive is the final fidelity with respect to small changes in the optimal control field?
To demonstrate the robustness of the optimized control field obtained using global optimization methods, we depict in Fig.~\ref{fig:opt-var} the comparisons between optimized control fields and the corresponding modification for the durations $T\!=\!2.6$ and $T\!=\!5$. The modified optimized control fields are obtained by mixing the optimized control field and linear ramping with different values of weight $w\in[0.1,\,0.9]$, i.e., $g(t)=w g_{\mathrm{Lin}}(t)+(1-w) g_{\mathrm{Opt}}(t)$. As shown in the top panel in Fig.~\ref{fig:opt-var}, the differences between the fidelity of the optimized control field $F_\mathrm{o}$  and that of the modification $F_\mathrm{w}$ increases as the weight of linear ramping grows (Note that $F_\mathrm{w} <F_\mathrm{o}$). For a special case where the weight is $10\%$, as shown in bottom panel in Fig.~\ref{fig:opt-var}, the differences between the fidelities are of order $10^{-3}$ for $T\!=\!2.6$ and $T\!=\!5$. It means that the optimized control fields obtained are robust to the imperfection or external noise which is naturally present in experiments. Thus, the scheme may be especially relevant in future experiments.

\section{Discussion}\label{sec5}
As two of the most promising evolutionary algorithms, the CMA-ES method and the SaDE method outperform other evolutionary computational algorithms and local optimization algorithms for high-dimensional optimization problems in certain quantum systems~\cite{shir2011,hansen2006,das2011}. Both algorithms, however, have its own advantages and disadvantages, and the preferences may vary from one case to another. The CMA-ES method is quasi parameter-free, while the SaDE method requires more initial parameters to be determined by the user. For the SaDE method, as mentioned in Sec.~\ref{sec3}, the population size $N_p$ is fixed to be $5D$ which is suggested as a lower bound and tested in great detail in the field of computational science (but not as extensively for physically motivated quantum systems). For the CMA-ES method, there is no guide for choosing the value of $N_p$, thus we choose $N_p=60$ which is large enough to guarantee that the fidelity is saturated. Therefore, in general, the population size of SaDE is larger than that of CMA-ES, especially in high-dimension parameter space, thus the computational time of SaDE is longer than CMA-ES. As for the convergence rate, in general, CMA-ES converges faster than SaDE. The slow convergence of SaDE is depicted in Fig.~\ref{fig:DE-ES} (bottom right) where the width of the staircase indicates the stagnation of the SaDE method. Note, however, if the $N_c$ of SaDE is the same as that of CMA-ES, the final fidelity obtained using the SaDE method is generally larger than that using the CMA-ES method, though the computational time of SaDE is much longer than CMA-ES. For instance, suppose that ($N_c=15, N_p=5D$) is taken for the SaDE method, which is the same as CMA-ES, then the computational time of SaDE is approximately three times larger than that of CMA-ES.

The calculations are performed in parallel using \textsc{Matlab} R2017a on cluster (Intel Xeon E5-2680 CPU with 28 cores and 251 GB RAM). Take CMA-ES for instance, the computational time of the CMA-ES method over 500 iterations is about 31h for $T=1.5$ and 71h for $T=2.5$.
For the same process duration $T$ and number of iterations, the computational time ratio of CMA-ES to SaDE is roughly $60:75$, which is the ratio of population size. This is because the maximal number of cores in the cluster is 28. If the number of cores is larger than the population size, then the computational time ratio of CMA-ES to SaDE is approximately 1:1.

\section{Conclusion}
We have given first numerical estimates of the quantum speed limit for three different few-fermion systems confined in a one-dimensional harmonic trap using the CMA-ES and the SaDE methods, and shown that the shortest duration obtained employing optimized, nonadiabatic processes is much faster than in the case of linear ramping and exponential ramping. One can achieve at least double speed-up in obtaining the target three-body ground state by using our optimized approach compared to the exponential ramping (see Fig.~\ref{fig:QSL_elo}). Since the Hilbert space increases greatly with the number of particles, the speed-up might increase even further for the systems with more than three particles. We observed that for durations shorter than the estimate of the quantum speed limit the optimized fields are of oscillation type, while for longer times, the optimized fields do not change drastically in time, which is analogous to the linear ramping and the exponential ramping. We have compared the performance of the CMA-ES and the SaDE methods, and found that the performance of CMA-ES is better than SaDE in terms of the best infidelity and convergence rate for the parameters considered in this paper. In addition, we have explained the advantages and disadvantages of the CMA-ES method, as well as the SaDE method, and the preference varies from one case to another. Moreover, we also demonstrated the robustness of the optimized control field to minor variation.
This stability of the above scheme on small variations let us believe that the obtained optimized fields are not just a purely numerical prediction, but can be useful in the noisy laboratory environment.

The present work shows the encouraging result that control theory can be used to obtain a significant speed up in producing a target state with the same symmetry as the initial state. As a next step, it would be very interesting to investigate how one can design control protocols to produce any of the low energy states in the spectrum with high fidelity starting from the ground state of the noninteracting system.

\begin{acknowledgments}
We would like to thank Nikolaj T. Zinner for discussions. This work has in part been supported by the Villum Foundation. DP and TS acknowledge support from the (Polish) National Science Center Grants No. 2016/21/N/ST2/03315 (DP) and 2016/22/E/ST2/00555 (TS). JS acknowledges support from ERC.
XL thanks the Max Planck Institute for the Physics of Complex Systems for hospitality during visits to the institute.

\end{acknowledgments}

\end{document}